%% file: iclr2021_workshop.tex
\documentclass{article}
\usepackage{iclr2021_workshop,times}

\input{math_commands.tex}

\usepackage{hyperref}
\usepackage{url}

\title{
\centering
ICLR 2021 Challenge for 

Computational Geometry \& Topology:

Design and Results}

\author{Nina Miolane\** \\
UC Santa Barbara, USA\\
\texttt{ninamiolane@ucsb.edu} \\
\And
Matteo Caorsi\** \\
L2F SA, Switzerland \\
\texttt{matteocao@gmail.com}
\And
Umberto Lupo\** \\
EPFL, Switzerland \\
\texttt{umberto.lupo@epfl.ch}
\And
Marius Guerard\** \\
Geomstats, USA \\
\And
Nicolas Guigui\** \\
Inria Sophia-Antipolis, France \\
\And
Johan Mathe\** \\
Atmo, USA \\
\And
Yann Cabanes\** \\
Institute of Mathematics of Bordeaux and Thales, France \\
\And
Wojciech Reise\** \\
Inria Paris-Saclay, France \\
\And
Thomas Davies\**\** \\
University of Southampton, UK \\
\And
Ant{ó}nio Leit{ã}o\**\** \\
NOVA IMS, Portugal \\
\And
Somesh Mohapatra\**\** \\
MIT, USA \\
\And
Saiteja Utpala\**\** \\
Mastercard, India \\
\And
Shailja Shailja\**\** \\
UC Santa Barbara, USA\\
\And
Gabriele Corso\**\** \\
Cambridge, UK\\
\And
Guoxi Liu\**\** \\
Clemson University, USA \\
\And
Federico Iuricich\**\** \\
Clemson University, USA \\
\And
Andrei Manolache\**\** \\
Bitdefender and University of Bucharest, Romania \\
\And
Mihaela Nistor\**\**, Matei Bejan\**\**, Armand Mihai Nicolicioiu\**\**, Bogdan-Alexandru Luchian\**\**\\
University of Bucharest, Romania \\
\And
Mihai-Sorin Stupariu\**\** \\
University of Bucharest, Romania \\
\And
Florent Michel\**\**, Khanh Dao Duc\**\** \\
University of British Columbia, Canada \\
\And
Bilal Abdulrahman\**\** \\
CUNY Graduate Center, USA \\
\And
Maxim Beketov\**\**\\
HSE University, Russia\\
\And
Elodie Maignant\**\** \\
Inria and ENS Paris-Saclay, France \\
\And
Zhiyuan Liu\**\** \\
University of North Carolina, USA \\
\And
Marek \v{C}ern\'{y}\**\** \\
Charles University, Czech Republic \\
\And
Martin Bauw, Santiago Velasco-Forero, Jesus Angulo\**\** \\
Mines ParisTech, CMM, PSL Research University, France \\
\And
Yanan Long\**\** \\
University of Chicago, USA \\
\AND
   \\
\vspace{-3mm}
\** : Organizers and external jury; \**\** : Participants
}

\iclrfinalcopy % Uncomment for camera-ready version, but NOT for submission.
\begin{document}

\maketitle

\begin{abstract}
    This paper presents the computational challenge on differential geometry and topology that happened within the ICLR 2021 workshop ``Geometric and Topological Representation Learning". The
    competition asked participants to provide creative contributions to the fields of computational geometry and topology through the open-source repositories \texttt{Geomstats} and \texttt{Giotto-TDA}. The challenge attracted 16 teams in its two month duration. This paper describes the design of the challenge and
    summarizes its main findings. 
    
    \textbf{Code:} \url{https://github.com/geomstats/challenge-iclr-2021}.
\end{abstract}

\section{Introduction}

%For example, Liu et al. \citep{Liu2019} present the successes and struggles encountered when trying to enable computational reproducibility of 14 articles submitted to a journal special issue. The authors state: ``Only 2 of the 14 manuscripts that were sub- mitted to the special issue were computationally reproducible initially, even though we had access to code provided by the authors. After an intensive revision process, 7 of the 12 accepted manuscripts eventually became computationally reproducible".

The motivation behind the organization of the ICLR 2021 Computational Geometry and Topology challenge was two-folds: first, to push forward the fields of computational differential geometry and topology; and second, to foster \textit{reproducible research} in mathematics by encouraging the use, development and maintenance of open-source repositories.

\paragraph{Reproducible Research} The \textit{reproducibility} of a (computational) experiment is widely regarded as a requirement to establish a scientific claim or to demonstrate the applicability of a technology. Different authors have suggested different levels of \textit{reproducibility} \citep{Dalle2013,Stodden2013,Schnell2018}. These levels range from \textit{repeatability} -- the ability of the same team to repeat the same experiment with the same methodological set up -- to the stronger notion of \textit{reproducibility} -- the ability of a different team to reproduce the results with a different methodological set up. In computational and mathematical sciences, such ``methodological setup" refers to the software environment, the data, and raw code. 

Despite many valuable initiatives to improve reproducibility in science, minimum standards are rarely met -- even in the mathematical sciences \citep{DavidRedish2018}. This so-called \textit{reproducibility crisis} has led researchers, funding agencies, politicians, and the wider audience to question the reliability of the scientific enterprise. Among other attempts to address the crisis, the workshop ``ICERM Workshop on Reproducibility in Computational and Experimental Mathematics (2012)" laid out several recommendations to the research community. In particular, the workshop committee claimed that ``appropriate tools" should be taught as standard operating procedure in relation to computational aspects of research \citep{Stodden2013}.

\paragraph{Computational notebooks} What are today's ``appropriate tools" in modern computational and mathematical sciences? The most commonly used tools may not be the most appropriate anymore. The publication process has barely evolved since the 17th century, as the research paper (or its electronic form) still represents the principal mean of diffusion for mathematical ideas. As James Sommers writes: ``Scientific results today are as often as not found with the help of computers. [...] And yet by far the most popular tool we have for communicating these results is the PDF - literally a simulation of a piece of paper. Maybe we can do better." 

Many new ``appropriate tools" are available to mathematical sciences to go beyond the sole diffusion format of the research paper. Among them, the computational notebook is an excellent candidate that can complement the traditional PDF paper \citep{Oakes2019}. Even a very theoretical mathematical paper could be completed by a computational notebook, that would: (i) use symbolic computation software to automatically check equations, (ii) leverage packages to check the veracity of a theorem on specific examples, (iii) provide interactive visualizations of special cases for the theoretical concepts exposed. This computational notebook could also be run automatically by the editor board of the corresponding journal, hence relieving some aspects of the review process and directly fostering reproducibility. 

\paragraph{Open-source packages} Computational notebooks would ideally heavily leverage a shared implementation of the core concepts of a given field of mathematics. This implementation would be materialized as a piece of open-source software, whose computations would be constantly checked by its community. As such, well-maintained open-source software and computational notebooks represent the foundational ``appropriate tools" that can foster reproducibility in mathematical research and with it, improve the efficiency and reliability of the research enterprise. Many open-source packages have made code available to foster reproducible research in their respective fields of mathematics. 

In the field of differential geometry, we find
\texttt{Pygeometry} \citep{Censi2012PyGeometry:Manifolds.}, \texttt{Manopt} \citep{Boumal2013a}, \texttt{Pyquaternion} \citep{Wynn2014PyQuaternions:Quaternions}, \texttt{Pyriemann} \citep{Barachant2015PyRiemann:Interface}, \texttt{PyManopt} \citep{Townsend2016}, \texttt{TheanoGeometry} \citep{Kuhnel2017ComputationalTheano}, \texttt{Geoopt} \citep{Kochurov2019Geoopt:Optim}, \texttt{Geomstats} \citep{Miolane2020}, the \texttt{SageManifold} project within the package \texttt{SageMath} \citep{Sagemath2020}, \texttt{Tensorflow Manopt} \citep{smirnov2021tensorflow}, and \texttt{JuliaManifolds} \citep{JuliaManifolds}, among others. In the field of topology, we find \texttt{Perseus} \citep{Perseus2012}, \texttt{Dipha} \citep{Dipha2014}, \texttt{Javaplex} \citep{Javaplex2014}, \texttt{TDA} \citep{fasy2015introduction}, \texttt{Dionysus} \citep{Dionysus}, \texttt{Eireen} \citep{Eireen2016}, \texttt{PHAT} \citep{PHAT2017}, the Topology ToolKit \texttt{TTK} \citep{TTK2018}, \texttt{RedHom} \citep{RedHom2019}, \texttt{Scikit-tda} \citep{scikittda2019}, \texttt{Giotto-TDA} \citep{tauzin2020giottotda}, \texttt{HomCloud} \citep{HomCloud2020}, \texttt{Diamorse} \citep{Diamorse2020}, \texttt{Gudhi} \citep{Gudhi2021}, \texttt{GDA-public} \citep{GDA-public}, and \texttt{Ripser} \citep{1908.02518}, to cite a few.

Despite the existence of these packages, computational notebooks do not always accompany the submission of a mathematical research paper. Furthermore, recruiting maintainers to ensure the validity of the code on these platforms is often difficult. Both issues can be explained by a lack of incentives in the associated scientific communities.

\paragraph{Incentives} Computational notebooks and open-source software -- such as the ones referenced in the previous paragraph -- are gaining popularity in several fields of research, for example within machine learning communities. However, they might be still under-used in the mathematical sciences for three main reasons. First, traditional mathematical training rarely introduce notebooks and software engineering as part of the curriculum. In differential geometry for example, textbooks may lack coding exercises or an associated interactive library. As a result, mathematicians do not necessarily master the tools available to use or write code associated with their findings. Second, many fields of mathematics lack a reference platform, such as a designated software, where researchers can share their computations and together contribute to their field. Third, there are only few incentives that motivate junior researchers in the mathematical sciences to learn good practices. The ``publish-or-perish" pressure can make it difficult for junior researchers to consider taking additional time to (learn to) implement and share their results. As a consequence, and specifically in differential geometry and topology, it can be challenging to reproduce results, even if they were produced by the same team.

\paragraph{Computational Geometry and Topology Challenge} The ICLR 2021 Computational Geometry and Topology Challenge aimed to address these issues by encouraging researchers to delve into open-source implementations of differential geometry and topology. The participants were asked to create computational notebooks using the open-source software \texttt{Geomstats} \citep{Miolane2020} and \texttt{Giotto-TDA} \citep{tauzin2020giottotda}. The goal was to showcase some of the aforementioned ``appropriate tools" for modern research in the mathematical sciences. The participants of the challenge were rewarded by the publication of the present paper and with prizes for the three winning teams.

\paragraph{Outline and contributions} The remainder of this paper is organized as follows. Section~\ref{sec:setup} introduces the setup and guidelines of the challenge. Section~\ref{sec:results} summarizes the submissions to the challenge. Section~\ref{sec:features} presents the main features used by the participants within the packages \texttt{Geomstats} and \texttt{Giotto-TDA}. Section~\ref{sec:limits} presents the limitations of the packages as reported by the participants. Section~\ref{sec:prop} provides a list of new features proposed by the participants that aim to enhance current implementations of computational geometry and topology. Section~\ref{sec:rank} describes the challenge's evaluation process and gives the final ranking of the submissions to the challenge.

\section{Setup of the challenge}\label{sec:setup}

The challenge was held in conjunction with the workshop ``Geometric and Topological Representation Learning" of the International Conference on Learning Representations (ICLR) 2021.

% as differential geometric and topological approaches are gaining popularity in machine learning and data science. 

%Modern real-world data typically belong to high dimensional ambient measurement spaces. Yet, they often exhibit low dimensional intrinsic structures that can be uncovered by geometric or topological methods. As a result, recent years have seen significant interest and progress in geometric and topological approaches to representation learning. Its goal was to acknowledge and reward excellent implementations of methods arising in this field. 

\paragraph{Guidelines} The participants were asked to submit a \texttt{Jupyter Notebook} to provide ``the best data analysis, computational method, or numerical experiment relying on state-of-the-art geometric and topological Python packages": \texttt{Geomstats} and \texttt{Giotto-TDA}. The participants submitted their \texttt{Jupyter Notebooks} via Pull Requests (PR) to the GitHub repository of the challenge \footnote{\url{https://github.com/geomstats/challenge-iclr-2021}}. Teams were accepted and there was no restriction on the number of team members. The current principal developers of \texttt{Geomstats} and \texttt{Giotto-TDA}, i.e. the co-authors of the published papers \citep{Miolane2020, tauzin2020giottotda}, were not allowed to participate.

Each submission was requested to respect the following structure: (i) Introduction and motivation, (ii) Analysis/Experiment, (iii) Benchmark, (iv) Limitations and perspectives. Guidelines were also giving examples of possible submissions:
\begin{itemize}
     \setlength\itemsep{-0.2em}
    \item Data analysis with geometric and topological methods,
    \item Implementation of a research paper with \texttt{Geomstats}/\texttt{Giotto-TDA}
    \item Implementation of a feature to merge into \texttt{Geomstats}/\texttt{Giotto-TDA} codebases,
    \item Implementation of visualization tools for \texttt{Geomstats}/\texttt{Giotto-TDA},
    \item Benchmarking/profiling on geometric and topological methods against other methods for a public dataset. 
\end{itemize}
This list was completed by the \texttt{submission-example-*} folders on the GitHub repository, to help participants understand the packages and design their submission.

\paragraph{Evaluation criterion: fostering creativity} The evaluation criterion was: ``how does the submission help push forward the fields of computational geometry and topology?". The submissions were ranked according to this evaluation criterion, through a voting procedure relying on the Condorcet method, see Section~\ref{sec:rank}. 

The choice of this evaluation criterion was motivated by several reasons. First, the criterion did not require participants to submit novel research. The main focus was the implementation, which could for instance be the reproduction of published research. Such a criterion can allow the participants to focus on producing clean code and to provide a hands-on explanation of the mathematical concepts at hand.

This criterion also did not bias participants towards showcasing ``positive results" such as a new method beating the state of the art. ``Negative results" were considered just as valuable as positive results. In particular, submissions criticizing the available packages, or showing examples where geometric and topological representations did not help the analysis were also significantly rewarded.

Lastly, such criterion encouraged participants to be generally creative. Most machine learning challenges are conducted by ranking the participants according to a quantitative metric on a test dataset. This can induce biases in the contributions of the participants, since methods that do not perform on that specific metric are not rewarded. While they have many other advantages, such criteria may hide interesting research. In contrast, our evaluation criterion, relying on a voting system through the Condorcet method, was meant to also reward creative submissions.

\paragraph{Software engineering practices} The participants were also encouraged to use software engineering best practices. Their code should be compatible with Python 3.8 and make an effort to respect the Python style guide PEP8. The \texttt{Jupyter notebooks} were automatically tested when a Pull Request was submitted and the tests were required to pass. If a dataset was used, the dataset had to be public and referenced. Participants could raise GitHub issues and/or request help or guidance at any time through \texttt{Geomstats} slack and \texttt{Giotto-TDA} slack. The help/guidance was be provided modulo availability of the maintainers.

\section{Submissions to the Challenge}\label{sec:results}

Sixteen teams submitted code before the deadline and participated in the challenge. This section provides a summary of their submissions.

\paragraph{Noise Invariant Topological Features}

This submission analyzes data topological structure whilst being robust to various data corruptions. Examples of perturbed data are noisy point clouds, photos taken from different views, or dynamic modeling. This submission showcases the pipeline for extracting Perturbed Topological Signatures (PTS) by using \texttt{Geomstats} and \texttt{Giotto-TDA} \citep{Som2018}. The topological properties are studied by using distance metrics and kernels defined on the Stiefel and Grassmann manifolds \citep{Edelman1999,Hamm2008,Som2018}. Experiments are performed on three datasets: SHREC 2010 \citep{SHREC2010}, Princeton COS429 \citep{COS429}, and MNIST \citep{MNIST2010}.

\paragraph{Estimators of Means of Symmetric Positive Matrices}

This submission investigates estimators of means of Symmetric Positive Definite (SPD) matrices. In the first notebook, the efficiency of the recursive estimation of the Karcher mean \citep{Ho2013}, and of the K-means algorithm relying on it, are benchmarked and improved. In the second notebook, the Shrinkage Estimator is implemented \citep{Yang2020}, and the notebook shows how it improves on the maximum likelihood estimator. Experiments rely on synthetic datasets on the manifold of SPD matrices using sampling methods on manifolds \citep{Schwartzman2016}.

\paragraph{Visualization of Kendall Shape Spaces for Triangles}

This submission introduces visualization methods for Kendall shape spaces of triangles. An object's shape can be described by locating relevant points on it, called landmarks \citep{Dryden2016}. The Kendall shape space of triangles in dimension $m=2, 3$ is the space of triangles quotiented by the group of rotations, translations and dilatations of $\mathbb{R}^m$ \citep{Kendall1984,Kendall1993}. This submission presents two new visualization methods of these Kendall shape spaces, demonstrates their use and compares them with an alternative visualization method for this dataset: the non-exact visualization of multidimensional scaling (MDS). The experiments are performed on synthetic data of triangles in $\mathbb{R}^m$ for $m=2, 3$, and on the dataset of optic nerve heads shapes from \texttt{Geomstats}' \texttt{datasets} module.

\paragraph{Map your Topology to Different Geometries}

This submission implements a method to map a set of points from one geometry of choice onto another while preserving the topology. In the context of this notebook, a ``geometry" refers to a Riemannian manifold such as Euclidean space, Hyperbolic space, Hypersphere, manifold of Symmetric Positive Definite (SPD) matrices, among others. The method uses gradient descent on Riemannian manifolds with a loss function introduced in \citep{Moor2020} that has been used in Deep Learning \citep{Moor2020, Chiappa2020}. Experiments are run on synthetic data generated on the Euclidean plane, the sphere and the Poincare ball.

\paragraph{Naive Image Anomaly Detection on Fashion MNIST}

This submission evaluates the possibility to achieve anomaly detection (AD) in image databases with naive distances to centroids and norms using Euclidean and Riemannian representations. The notebook considers simple AD setups where the objective is to discriminate between two classes of the Fashion MNIST dataset \citep{fMNIST2017}. A general approach to embed images into the space of covariance matrices is introduced based on \citep{calvo1990distance}. The best performances are achieved by the method relying on the norm of the negated geodesic principal component analysis (PCA) with the Fr\'echet mean as PCA base point \citep{Rippel2020}, using the Log-Euclidean Riemannian metric.

\paragraph{Shape Analysis of Bone Cancer Cells}

This submission studies osteosarcoma (bone cancer) cells and the impact of drug treatment on their morphological shapes. The analysis uses cell images obtained from fluorescence microscopy. The corresponding dataset has been added into the \texttt{Geomstats}' module \texttt{datasets} by the participants. Cell shapes are modelled as discrete (open) curves. The submission uses the Riemannian elastic metric on discrete curves to compare cell shapes \citep{Jermyn2011}. The biological assumption is that such measures of irregularity and spreading of cells allow accurate classification and discrimination between cancer cell lines treated with different drugs \citep{Alizadeh2019}. The submission studies to which extent this Riemannian metric can detect how the cell shape is associated with the response to treatment.

\paragraph{Repurposing Peptide Inhibitors for SARS-Cov-2 Spike Protein}

This submission develops an approach combining physico-chemical parameter analysis and topological featurization to train robust one-class classifiers to predict protein-protein interactions (PPIs). PPIs form the molecular basis of processes that equally sustain life and drive development of disease, such as SARS-Cov-2. Peptides have garnered therapeutic interest due to their potential to disrupt clinically-relevant PPIs, apart from synthetic accessibility and better targeting modalities \citep{PTsomaia2015, Nina2020, Schissel2020}. The submission uses the top-performing model to screen the peptides in the current dataset against SARS-Cov-2 receptor binding domain protein. The Peptide Binding DataBase (PepBDB) is used for model training \citep{Wen2018}.

\paragraph{Shape analysis with skeletal models and Principal Nested Spheres}

This submission considers anatomical shape analysis with skeletal representations (s-reps) \citep{Liu2021} and Principal Nested Spheres (PNS) \citep{Jung2012, Kim2020}. The s-rep of a given shape consists of the shape's skeleton and two functions defined on the skeleton: a radial vector field and a radius function. PNS is a manifold learning method that addresses the non-Euclidean properties of shape data. PNS fits a \textit{hierarchy} of submanifolds -- subspheres -- to some input data. The notebook applies this method to s-reps of toy data and to the classification problem of the hand skeleton shape dataset available in \texttt{Geomstats}' \texttt{datasets} module, comparing s-reps and PNS to Euclidean and Riemannian alternatives from the literature. The best classification performance is obtained by using the Kendall Riemmanian metric \citep{Kendall1993} on the hand skeleton shapes.

\paragraph{Riemannian mean-shift algorithm}

This submission implements a Riemannian version of the mean-shift algorithm \citep{Subbarao09,Caisero2012}. Classic (Euclidean) mean shift works by sliding a window (a ball whose radius is called ``bandwidth") over the dataset, iteratively adjusting the center of the window until convergence to the estimated mode of the data. Mean shift is used for clustering, with several advantages over K-means. This notebook implements the method and shows its applicability on toy datasets on the sphere and hyperbolic plane.

\paragraph{Intrinsic Disease Maps Using Persistent Cohomology}

This submission uses persistent cohomology to investigate and visualize two infectious disease progression datasets: physiological data on Malaria in mice \citep{Cumnock2018} and humans \citep{Torres2016}, and data on Hepatitis C in humans \citep{Rosenberg2018}. The submission reiterates the work of \citep{Amin2021} and computes circular coordinates using the methodology introduced in \citep{deSilva2009}. The generated circular coordinate function provides an intrinsic disease phase coordinate that maps out the disease progression in the full data space.

\paragraph{Neural Sequence Distance Embeddings}

This submission presents Neural Sequence Distance Embeddings (\texttt{NeuroSEED}), a general framework to embed biological sequences in geometric vector spaces that reflect their evolutionary distance. The notebook illustrates the effectiveness of the hyperbolic space that captures the hierarchical structure and provides an average 38\% reduction in embedding RMSE against the best competing geometry. The capacity of the framework and the significance of these improvements are then demonstrated devising supervised and unsupervised \texttt{NeuroSEED} approaches to multiple core tasks in bioinformatics. Benchmarked with common baselines, the proposed approaches display significant accuracy and/or runtime improvements on real-world datasets \citep{clemente2015microbiome, zheng2019sense}. 

\paragraph{Analyzing Representative Cycles for Persistent Homology}

This submission aims to simplify the use of cycles for the analysis of persistent homology. The persistence diagram is often the only representation that software packages for TDA provide to visualize persistent homology information. Visualizing where each homology class appeared in the domain space can be very challenging for a user. This submission provides a collection of functions to simplify the interactive visualization and analysis of the homology class by enriching the information contained in the persistence diagram with cycles. Cycles are computed with an external library \cite{PersCycles2020} which uses Discrete Morse theory \cite{Robins2011,Mischaikow2013} to achieve scalability. The notebook demonstrates the use of cycles on a subset of the MNIST dataset \citep{MNIST2010} and provides an overview of applications using the direct visualization of cycles for exploratory data analysis.

\paragraph{Investigating CNN weights with Giotto Vectorization}

This submission provides a topological analysis of convolutional neural networks (CNN) weights. Transforming each layer to a new auxiliary space predicts network properties on a non-trivial supervised classification task. The notebook uses the Small CNN Zoo dataset \citep{Unterthiner2021}, shows how to compute the persistence diagrams of the auxiliary space, vectorizes the diagrams into Silhouettes \citep{Chazal2014}, and finally runs several regression and classification experiments. The results are particularly encouraging in terms of anomaly detection.

\paragraph{Brain Connectomes Comparison using Geodesic Distances}

This submission investigates the performance of geodesic distances on manifolds to assess brain connectome similarity between pairs of twins in terms of their structural networks at different network resolutions. The notebook uses the brain structural connectomes of 412 human subjects in five different resolutions and two edge weights \citep{Kerepesi2016}. The notebook investigates the performance of geodesic distances on manifolds and compares them with Euclidean distances within a Wilcoxon rank sum non-parametric test \citep{Cuzick1985}.

\paragraph{Fuzzy c-Means Clustering for Persistence Diagrams and Riemannian Manifolds}

This submission implements Fuzzy c-Means clustering for persistence diagrams and Riemannian manifolds \citep{Davies2020}. Many real world problems are fuzzy; that is, data points can have partial membership to several clusters, rather than a single ``hard" labelling to only one cluster \citep{Campello2007}. The notebook describes the fuzzy c-means algorithm, highlights the convergence results, and demonstrates fuzzy clustering on two simple datasets.

\paragraph{Reweighting Vectors for Graph Convolutional Neural Networks via Poincaré Embedding and Persistence Images}

This submission demonstrates how to incorporate local graph topological properties (e.g.\ connected components, cycles) into persistence enhanced graph neural networks (GNN) \citep{Zhao2020} for graph and node classification tasks. The notebook converts unweighted graphs to weighted graphs by embedding them using the Poincar\'e ball model and using the resulting Riemannian distances as the weights. Then, the persistence images of resulting weighted graphs are computed and the resulting matrix is used to re-weight the GNN weights for enhanced performances.

\section{Features used in the packages}\label{sec:features}

This section presents the features used in both packages and the limitations outlined by the participants. The numbers in parentheses refer to the number of submissions in which a given feature has been used.

\paragraph{Features used in \texttt{Geomstats}}

The differential geometric structures used in the submissions are the following: \texttt{Hypersphere} ($\times 4$), Kendall's \texttt{PreShapeSpace} ($\times 3$) with associated \texttt{KendallShapeMetric} ($\times 3$), the space of symmetric positive definite matrices \texttt{SPDMatrices} ($\times 2$) with associated Bures-Wasserstein metric \texttt{SPDMetricBuresWasserstein} ($\times 2$), or alternative Riemannian metrics such as the \texttt{SPDMetricAffine} ($\times 1$) and the \texttt{SPDMetricLogEuclidean} ($\times 1$), the Lie group of rotations \texttt{SpecialOrthogonal} ($\times 2$), the hyperbolic space in its \texttt{PoincareBall} ($\times 2$) representation and associated \texttt{PoincareBallMetric}  ($\times 1$), the space of \texttt{Matrices} ($\times 1$), the general class of \texttt{RiemannianMetric} ($\times 1$), the manifold of \texttt{DiscreteCurves} ($\times 1$) with associated square-root velocity metric \texttt{SRVMetric} ($\times 1$), the \texttt{Grassmanian} ($\times 1$) with the \texttt{GrassmanianCanonicalMetric} ($\times 1$), \texttt{Stiefel} ($\times 1$) with the \texttt{StiefelCanonicalMetric} ($\times 1$). The algorithms of geometric statistics that have been used are: the estimation of the \texttt{FrechetMean} ($\times 6$), the tangent principal component analysis \texttt{TangentPCA} ($\times 3$), the \texttt{RiemannianKMeans} and the Riemannian version of the \texttt{KNearest} \texttt{NeighbordsClassifier}. In terms of differential geometric datasets, the cell shapes dataset, hand skeleton shape dataset and the optical nerve shape dataset were used. The visualization module was also used through its \texttt{Sphere} ($\times 3$), \texttt{PoincareDisk} ($\times 2$), \texttt{KendallDisk} ($\times 1$), \texttt{KendallSphere} ($\times 1$). The main features that have not been used are the mathematical structures and functions related to information geometry, such as the manifolds of Beta distributions, Dirichlet distributions and Normal distributions; and the more involved learning procedures such as the \texttt{expectation\_maximization} or the \texttt{KalmanFilter} on Lie groups.

\paragraph{Features used in \texttt{Giotto-TDA}}

The topological features used in the submissions are the following. \texttt{VietorisRipsPersistence} ($\times 6$) was the most used class for computing persistent homology, followed by \texttt{CubicalPersistence} ($\times 1$). In the \texttt{diagrams} module, the computation of distance matrices between persistence diagrams via \texttt{PairwiseDistance} ($\times 3$) was used but not preferred to the vectorisation of persistence diagrams via \texttt{PersistenceImage} ($\times 2$), \texttt{PersistenceEntropy} ($\times 1$), \texttt{NumberOfPoints} ($\times 1$), \texttt{Amplitude} ($\times 1$) and \texttt{Silhouette} ($\times 1$). Additionally, \texttt{Scaler} ($\times 1$), from the same module, was used.  A few pre-processing utilities for images were explored, namely \texttt{Binarizer} ($\times 2$) and the following classes for creating filtrations from 2D or 3D greyscale images: \texttt{RadialFiltration} ($\times 2$), \texttt{DilationFiltration} ($\times 1$), \texttt{ErosionFiltration} ($\times 1$) and \texttt{DensityFiltration} ($\times 1$). The visualization module \texttt{plotting} was used through its functions \texttt{plot\_diagram} ($\times 3$) and \texttt{plot\_point\_cloud} ($\times 2$). Among the main modules that have not been used, we find the modules related to time-series and curves.

\section{Limitations of the packages}\label{sec:limits}

Participants were asked to report on the limitations of the packages. This section provides a summary of their findings.

\paragraph{Limitations of \texttt{Geomstats}} First, some participants reported bugs. For example, there was an issue to project points from the \texttt{Grassmannian} or the \texttt{Stiefel} manifold to the tangent space using default canonical metrics or the participants' custom metrics. Both \texttt{preprocessing.ToTangentSpace} and \texttt{Grassmannian.to\_tangent} failed with the same issue. A similar problem was encountered while trying to project points from the manifold of \texttt{DiscreteCurves} on its tangent space at the \texttt{FrechetMean} using the square-root velocity metric \texttt{SRVMetric}. In this case, the implementation of the \texttt{FrechetMean} itself was failing.

Second, participants did not find some implementations in the package or struggled to understand the existing code. For example, a participant reported that the metric for the \texttt{HyperbolicSpace} was missing, and another tried to use the abstract class \texttt{RiemannianMetric} without providing a definition of inner-product or a metric matrix. Another participant would have liked to use product manifold and product metrics but did not realize that it was implemented in the library. Other participants wanted to use several backends but could not find the way to use both in one script: \texttt{os.environ["GEOMSTATS\_BACKEND"] = "pytorch"; importlib.reload(geomstats.backend)}. 

%The participant tried to fix it, without success due to the lack of documentation and understanding of the package.

These issues come from a lack of completeness of the current documentation of the package, misleading error messages and possible erroneous existing documentation. Other participants did report several problems with the current documentation, which could be improved with more detailed descriptions and an index with short summaries. There are classes such as the \texttt{PoincareBall} that are found in tutorials, but not found in the documentation website. 

Lastly, participants reported the lack of integration between the modules related to graph and hyperbolic spaces in \texttt{Geomstats}, and the formalism of modern libraries using graphs. In this case, a refactoring is needed to allow a better integration between geometric statistics through \texttt{Geomstats} and packages of geometric learning such as \texttt{pytorch-geometric} \citep{PytorchGeom2019}, \texttt{networkx} \citep{NETWORKX} and \texttt{dgl} \citep{Dgl2019}.

\paragraph{Limitations of \texttt{Giotto-TDA}}  First, some participants reported bugs. For example, computing persistence images on persistence diagrams formed by a single persistence pair outputs an empty persistence image.

% Then, some participants struggled to find existing features. For example, one participant reported that it was not possible to compute the persistent homology of point clouds with different number of points, which along with the first point did not allow to compute the pairwise distance matrix of point clouds with different number of points. However, the \texttt{VietorisRipsPersistence} handles points clouds of different cardinalities and in different dimensions. As for \texttt{Geomstats}, a clearer documentation might be able to address this issue.

Some participants pointed out that the rigid input requirements for \texttt{PairwiseDistance} -- in particular, the fact that all persistence diagrams must formally have the same number of homology dimensions and of birth-death pairs in each homology dimension -- can be limiting in applications.  This indicates that a utility function for converting collections of persistence diagrams into a format accepted by \texttt{PairwiseDistance} should be added to the package. Possibly related to this, some participants surmised (incorrectly, in this case) that persistent homology transformers such as \texttt{VietorisRipsPersistence} cannot handle collections of point clouds of different cardinalities. This might indicate that the aforementioned tight requirements in \texttt{PairwiseDistance} are at odds with the more permissive character of many other components of \texttt{Giotto-tda}.

The next reported limitation came from the architecture of the package itself. Some participants reported that the package's high-level API did not allow for manipulations of attributes and usage of methods that are present in the objects that it runs underneath. For example, \texttt{VietorisRipsPersistence} runs using \texttt{Ripser} but does not allow one to access the co-cycles that are otherwise accessible using \texttt{Ripser}'s API directly.

Lastly, in terms of performance, some participants reported that the computational runtime for \texttt{PersistenceImage} was very irregular on their data in comparison to the performance of e.g.\ \texttt{Silhouette}.

\section{Proposed features for the packages}\label{sec:prop}

This section lists the features, suggested by the participants, that could be implemented in packages of computational geometry and topology such as \texttt{Geomstats} and \texttt{Giotto-TDA}. These are implementations that they would judge useful in order to push forwards the fields of computational geometry and topology.

\paragraph{Proposed features for \texttt{Geomstats}} As the Stiefel and the Grassmann manifolds are becoming popular in the Machine Learning and Computer Vision community, more geometric features on these manifolds (such as various metrics) could offer powerful tools for solving a wide range of learning tasks. 

In the same vein, symmetric positive definite (SPD) matrices are raising more and more interest in the same communities. While \texttt{Geomstats} has a module that processes time-series into SPD matrices, further SPD-dedicated preprocessing to SPD matrices for various data types would be helpful. Trainable SPD representations would also be of interest. For instance, some participants specifically asked for an implementation of SPD neural networks, the so-called second order neural networks.

Then, the module on the geometry of discrete curves could be improved in different ways. \texttt{Geomstats} only provides spaces of open curves, a restriction which is also not necessarily clear through reading the documentation only. Adding the implementation of the space of closed curves would be interesting. Furthermore, only the elastic metric on these spaces of curves is implemented, while generalizations of this metric exist that could be interesting. We note that these last two structures have been implemented in the library since. Lastly, when looking at shapes of curves, it is interesting to quotient out not only the reparameterization of the curve (done with the elastic metric), but also the rotations and translations of these curves. This is not obvious from the current documentation, and indications of this aspect, together with a recommendation to use the module of Kendall shape space for this, could be helpful.

Lastly, the visualization module could be further improved, by allowing more interactive visualizations and adding a visualization for the space of SPD matrices that could allow to visualize the differences between the different metrics on SPD matrices (of low dimensions). 

\paragraph{Proposed features for \texttt{Giotto-TDA}}

First, participants suggested to add an implementation that computes the pairwise distance matrix of diagrams with different number of homology groups in each dimension, without having to first perform some laborious manual padding. Some participants also offered to add tools to compute the cohomology persistence, and for example circular coordinates.

For machine learning applications, some participants suggested to use a tensor backend instead of NumPy, such as Tensorflow or PyTorch.

% We detected three improvement ideas during our experiments that can make the Giotto library easier to use.

% The option of selecting the number of bins in gtda.diagrams.Silhouette could be extended so that the user can input a fixed scale of the bins.
% 
% For machine learning applications, we suggest the use tensor backend instead of NumPy.

%######

%New estimators of mean that are faster. Possiblity of using Kmeans based on these different estimators. A sampling module.

%Develop of the vis module with interactive visualizations.
% A class that can be used to plot points in SPD space would be helpful in visualizing the geodesic distances.

% The first limitation that we faced is that Geomstats current implementation only deals with open curves. This restriction to open curves was not very clear when we first looked at the documentation. In the context of our application, having an implementation with closed curves would be quite relevant as cell boundaries are naturally closed.

% closed curves. other elastic metrics. quotient by rotations.

% Curves shapes:
% To compare curves, we also had to preprocess the data, to take into account the fact that Geomstats does not consider that cells are defined up to transformations (such as rotations). 

\section{Final ranking}\label{sec:rank}

This section provides the final ranking of the challenge's submissions. The Condorcet method was used to rank the submissions based on the single evaluation criterion: ``how does the submission help push forward the fields of computational geometry and topology?"

Each of the 16 teams had the opportunity to vote for the 3 best submissions. Each team received only one vote, even if there were several participants in the team. In addition, 8 external reviewers, chosen among \texttt{Geomstats} and \texttt{Giotto-TDA} core maintainers and all from different institutions, also voted for the 3 best submissions. The 3 preferences had to be all 3 be different: e.g. one could not select the same \texttt{Jupyter Notebook} for both first and second place. The submissions were anonymized, the votes remained secret, only the final ranking is published here. Ties are represented by bullet points in the ranking below.

\begin{enumerate}
    \setlength\itemsep{-0.2em}
    \item Noise Invariant topological features
    \item \begin{itemize}
        \setlength\itemsep{-0.2em}
        \item Estimators of Means of Symmetric Positive Matrices
        \item Visualization of Kendall Shape Spaces for Triangles
    \end{itemize}
    \item Neural Sequence Distance Embeddings
    \item Repurposing Peptide Inhibitors for SARS-Cov-2 Spike Protein
    \item \begin{itemize}
        \setlength\itemsep{-0.2em}
        \item Map your Topology to Different Geometries 
        \item Intrinsic Disease Maps Using Persistent Cohomology
    \end{itemize}
    \item Shape analysis with skeletal models and Principal Nested Spheres
    \item \begin{itemize}
        \setlength\itemsep{-0.2em}
        \item Riemannian mean-shift algorithm
        \item Fuzzy c-Means Clustering for Persistence Diagrams and Riemannian Manifolds
    \end{itemize}
    \item \begin{itemize}
        \setlength\itemsep{-0.2em}
        \item Naive Image Anomaly Detection on Fashion MNIST
        \item Shape Analysis of Bone Cancer Cells
        \item Brain Connectomes Comparison using Geodesic Distances
        \item Investigating CNN weights with Giotto Vectorization
    \end{itemize}
    \item Analyzing Representative Cycles for Persistent Homology
    \item Reweighting Vectors for Graph Convolutional Neural Networks via Poincaré Embedding and Persistence Images
\end{enumerate}

Regardless of this final ranking, we would like to stress that all the submissions were of very high quality. We warmly congratulate all the participants.

\subsubsection*{Author Contributions}

Nina Miolane, Matteo Corsi, Umberto Lupo, Marius Guerard and Nicolas Guigui led the organization of the challenge. Nina Miolane and Marius Guerard were responsible of the GitHub repository. Nina Miolane, Matteo Caorsi, Umberto Lupo, Marius Guerard, Nicolas Guigui, Johan Mathe, Yann Cabanes and Wojciech Reise were the external reviewers in the evaluation process. The remaining authors of this white paper were the participants of the challenge.

\subsubsection*{Acknowledgments}

The authors would like the thank the organizers of the ICLR 2021 workshop ``Geometrical and Topological Representation Learning" for their valuable support in the organization of the challenge and specifically Bastian Rieck for his availability and help.

\section{Conclusion}

This white paper presented the motivations behind the organization of the ``Computational Geometric and Topological Challenge" at the ICLR 2021 workshop ``Geometric and Topological Representation Learning", and summarized the findings from the participants' submissions.

\bibliography{iclr2021_workshop}
\bibliographystyle{iclr2021_workshop}
\end{document}

%% file: math_commands.tex
\usepackage{amsmath,amsfonts,bm}

\def\eqref#1{equation~\ref{#1}}
\def\1{\bm{1}}

\DeclareMathAlphabet{\mathsfit}{\encodingdefault}{\sfdefault}{m}{sl}
\SetMathAlphabet{\mathsfit}{bold}{\encodingdefault}{\sfdefault}{bx}{n}